\documentclass[prb,twocolumn,showpacs,preprintnumbers,amsmath,amssymb,floatfix]{revtex4-1}

\usepackage{graphicx,color,hyperref}

\begin{document}

\title{Electrons and phonons in single layers of hexagonal indium
  chalcogenides from {\it ab initio} calculations}

\author{V.\ Z\'{o}lyomi, N.\ D.\ Drummond, and V.\ I.\ Fal'ko}

\affiliation{Physics Department, Lancaster University, Lancaster LA1 4YB,
  United Kingdom}

\date{\today}

\begin{abstract} We use density functional theory to calculate the
electronic band structures, cohesive energies, phonon dispersions, and optical
absorption spectra of two-dimensional In$_2$X$_2$ crystals, where X is S, Se,
or Te.  We identify two crystalline phases ($\alpha$ and $\beta$) of
monolayers of hexagonal In$_2$X$_2$, and show that they are characterized by
different sets of Raman-active phonon modes.  We find that these materials are
indirect-band-gap semiconductors with a sombrero-shaped dispersion of holes
near the valence-band edge. The latter feature results in a Lifshitz
transition (a change in the Fermi-surface topology of hole-doped In$_2$X$_2$)
at hole concentrations $n_{\rm S}=6.86\times 10^{13}$ cm$^{-2}$, $n_{\rm
  Se}=6.20\times 10^{13}$ cm$^{-2}$, and $n_{\rm Te}=2.86\times 10^{13}$
cm$^{-2}$ for X=S, Se, and Te, respectively, for $\alpha$-In$_2$X$_2$ and
$n_{\rm S}=8.32\times 10^{13}$ cm$^{-2}$, $n_{\rm Se}=6.00\times 10^{13}$
cm$^{-2}$, and $n_{\rm Te}=8.14\times 10^{13}$ cm$^{-2}$ for
$\beta$-In$_2$X$_2$. \end{abstract}


\maketitle

\section{Introduction}

The discovery of graphene\cite{novoselov_2004,geim_2007} has triggered the
growth of a family of two-dimensional (2D) nanomaterials, including hexagonal
boron nitride,\cite{BN01,BN02}
silicene,\cite{Silicene01,Silicene02,Silicene03,DrummondSilicene}
germanane,\cite{Germanane} and a variety of transition metal
dichalcogenides.\cite{mos2exfol,KisA_2011_2,KisA_2011,WS2exp01,WS2exp02} These
materials are of great interest due to their potential applications in
optoelectronics.\cite{KisA_2011_2,AtacaC_2012,WS2exp01,Morpurgo} Recently we
discussed a new member of this family: atomically thin layers of hexagonal
gallium chalcogenides,\cite{ZolyomiGaX} which are indirect-band-gap
semiconductors with unusual, sombrero-shaped valence-band edges and optical
absorption spectra that are dominated by zone-edge transitions. In this work
we study closely related materials: 2D crystals of indium chalcogenides
(In$_2$X$_2$, where X is S, Se, or Te).

Chalcogenides of indium take several
forms,\cite{BarronBook,InSeLit01,InSeLit02,InSeLit03,InSeLit04} including
tetragonal, rhombohedral, cubic, monoclinic, and orthorhombic phases, as well
as the hexagonal structures on which we focus here. Indium selenide (InSe)
exists in a layered hexagonal structure in nature with an in-plane lattice
parameter of 4.05 {\AA} and a vertical lattice parameter of 16.93 {\AA}, and
has been proposed for use in ultrahigh-density electron-beam-based data
storage.\cite{GibsonGA_2005} Very recently, samples of few-layer hexagonal
InSe have been produced and their optical properties have been
studied.\cite{lei_2014,Mudd2014} Indium sulfide (InS) and indium telluride
(InTe) exhibit orthorhombic and tetragonal structures, respectively, but this
does not exclude the possibility of growing metastable hexagonal structures
(structural changes induced by annealing have been reported in transmission
electron microscopy of indium chalcogenide thin films\cite{FitzgeraldIOP}).
We have investigated whether monolayers of the hexagonal phase are stable in
any of these three materials.

\begin{figure}
\begin{center}
\includegraphics[clip,width=0.45\textwidth]{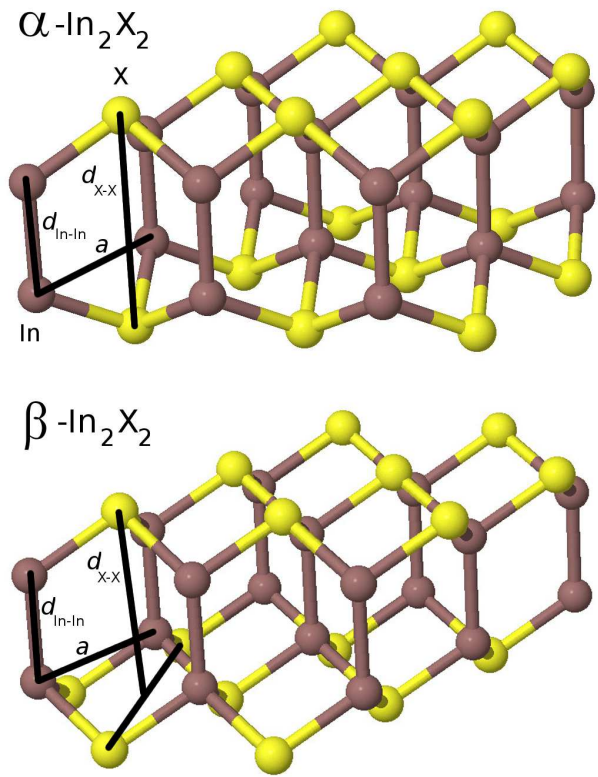}
\caption{(Color online) Structures of the $\alpha$ and $\beta$ polytypes of
  monolayer indium chalcogenides In$_2$X$_2$ (X=S, Se, or Te).  The parameters
  $a$, $d_{\rm{In}-\rm{In}}$, and $d_{\rm{X}-\rm{X}}$ are the lattice
  parameter, the In--In bond length, and the vertical distance between X
  atoms, respectively.
\label{fig:structure}}
\end{center}
\end{figure}

The structures of two stable or metastable polytypes of monolayer hexagonal
In$_2$X$_2$ identified in this work are shown in Fig.\ \ref{fig:structure}.
Viewed from above, a monolayer of $\alpha$-In$_2$X$_2$ forms a 2D honeycomb
lattice, with vertically aligned In$_2$ and X$_2$ pairs at the different
sublattice sites. Its point group is $D_{3h}$. The $sp$ orbitals of the In
atoms in each dimer are strongly hybridized, and each of the two In atoms is
bound to three neighboring chalcogens. The lattice structure of
$\beta$-In$_2$X$_2$ is depicted in the bottom panel of
Fig.\ \ref{fig:structure}, with one of the X layers shifted with respect to
the other, breaking the mirror symmetry of the original structure but
establishing inversion symmetry in its stead. The point group of
$\beta$-In$_2$X$_2$ is $D_{3d}$.  The lattice parameters calculated using
\textit{ab initio} density functional theory (DFT) for these two polytypes of
In$_2$X$_2$ are discussed in Sec.\ \ref{sec:results}, along with lattice
dynamics.  We find that the $\alpha$ and $\beta$ polytypes can be
distinguished by comparing optically active [infrared (IR) and Raman] phonon
spectra and that the band structures of $\alpha$-In$_2$X$_2$ crystals are very
similar to those of hexagonal Ga$_2$X$_2$ crystals.\cite{ZolyomiGaX} In
Secs.\ \ref{sec:alpha} and \ref{sec:beta} we report first-principles
calculations of the electronic band structures of $\alpha$-In$_2$X$_2$ and
$\beta$-In$_2$X$_2$.

Our DFT calculations were performed using the \textsc{castep}\cite{castep} and
\textsc{vasp}\cite{vasp} plane-wave-basis codes to calculate the structural
parameters of In$_2$X$_2$.  We used both the local density approximation (LDA)
and the Perdew-Burke-Ernzerhof\cite{pbe} (PBE) generalized gradient
approximation exchange-correlation functionals in our calculations. The same
functionals were used to calculate the electronic band structures, optical
absorption spectra, and phonon dispersion curves.  For the electronic band
structures we also used the screened Heyd-Scuseria-Ernzerhof 06 (HSE06) hybrid
functional\cite{hse} to compensate at least partially for the underestimation
of the band gap by the LDA and PBE functionals.  The HSE06 band structure
calculations used the geometry optimized using the PBE functional.  The
plane-wave cutoff energy used in our calculations was 600 eV\@. During the
geometry relaxations a $12\times 12$ Monkhorst-Pack \textbf{k}-point grid was
used, while band structures were obtained with a $24\times 24$ grid. The
optical absorption spectra were obtained with a very dense grid of $95\times
95$ \textbf{k} points.  The artificial out-of-plane periodicity of the
monolayer was set to 20 {\AA} in each case.  Phonon dispersion curves were
calculated in \textsc{vasp} using the method of finite displacements in a
$4\times 4$ supercell with $6 \times 6$ \textbf{k}-points, and in
\textsc{castep}\cite{castep2} using density functional perturbation theory
(DFPT)\@.  We also evaluated the infrared intensity and Raman intensity
tensors for the zone-center optical phonons in In$_2$X$_2$.  The DFPT
calculations used a plane-wave cutoff of 816 eV, a $31 \times 31$
Monkhorst--Pack grid, norm-conserving DFT pseudopotentials, and an artificial
periodicity of 15.9 {\AA}.

\section{Lattice structure and lattice dynamics of $\alpha$-In$_2$X$_2$ and
  $\beta$-In$_2$X$_2$ \label{sec:results}}

\subsection{Lattice structures}

Our geometry-optimization calculations show that the lattice parameters in
$\alpha$-In$_2$X$_2$ increase with the atomic number of the chalcogen atom X,
while the In--In bond lengths hardly change: see Table
\ref{table:parameters_summary}.  The bond lengths obtained with the PBE
functional are systematically larger than those optimized within the LDA, as
expected.\cite{FavotF_1999} As shown in Sec.\ \ref{sec:phonons}, we find all
three $\alpha$-In$_2$X$_2$ crystals to be dynamically stable.  The cohesive
energy $E_c$ is also shown in Table \ref{table:parameters_summary}.  This is
the energy of two isolated indium atoms plus the energy of two isolated
chalcogen atoms minus the energy per unit cell of the In$_2$X$_2$ layer.  We
have not included the zero-point phonon energy in the latter. The difference
between the LDA and PBE cohesive energies is significant; nevertheless, both
functionals predict the cohesive energy to be largest for In$_2$S$_2$ and
smallest for In$_2$Te$_2$.

\begingroup \squeezetable
\begin{table}
\caption{Structural parameters (as defined in Fig.\ \ref{fig:structure}) of
  monolayer $\alpha$-In$_2$X$_2$ (top) and $\beta$-In$_2$X$_2$ (bottom) from
  DFT calculations with the LDA and PBE exchange-correlation functionals. The
  static-lattice cohesive (atomization) energy $E_c$ is also shown, as is the
  phonon ZPE\@.
\label{table:parameters_summary}}

\begin{tabular}{lcccccccccc}
\hline \hline

\multicolumn{11}{c}{$\alpha$-In$_2$X$_2$} \\

& \multicolumn{2}{c}{$a$ ({\AA})} & \multicolumn{2}{c} {$d_{\rm{In}-\rm{In}}$
  ({\AA})} & \multicolumn{2}{c} {$d_{\rm{X}-\rm{X}}$ ({\AA})} &
\multicolumn{2}{c} {$E_c$ (eV/cell)} & \multicolumn{2}{c} {ZPE (eV/cell)} \\

\raisebox{1.5ex}[0pt]{X} & LDA & PBE & LDA & PBE & LDA & PBE & LDA & PBE & LDA
& PBE \\

\hline

S & $3.80$ & $3.92$ & $2.74$ & $2.83$ & $5.11$ & $5.18$ & $16.17$ & $13.85$ &
$0.135$ & $0.127$ \\

Se & $3.95$ & $4.09$ & $2.74$ & $2.83$ & $5.30$ & $5.38$ & $15.12$ & $12.87$ &
$0.097$ & $0.091$  \\

Te & $4.23$ & $4.38$ & $2.73$ & $2.82$ & $5.50$ & $5.60$ & $14.00$ & $11.87$ &
$0.080$ & $0.075$ \\

\hline \hline

\multicolumn{11}{c}{$\beta$-In$_2$X$_2$} \\

& \multicolumn{2}{c}{$a$ ({\AA})} & \multicolumn{2}{c} {$d_{\rm{In}-\rm{In}}$
  ({\AA})}  & \multicolumn{2}{c} {$d_{\rm{X}-\rm{X}}$ ({\AA})} &
\multicolumn{2}{c} {$E_c$ (eV/cell)} & \multicolumn{2}{c} {ZPE (eV/cell)} \\

\raisebox{1.5ex}[0pt]{X} & LDA & PBE & LDA & PBE & LDA & PBE & LDA & PBE & LDA
& PBE \\

\hline

S & $3.81$ & $3.93$ & $2.74$ & $2.83$ & $5.10$ & $5.17$ & $16.15$ & $13.84$ &
$0.135$ & $0.127$ \\

Se & $3.96$ & $4.09$ & $2.74$ & $2.82$ & $5.28$ & $5.37$ & $15.10$ & $12.86$ &
$0.097$ & $0.091$ \\

Te & $4.24$ & $4.39$ & $2.73$ & $2.82$ & $5.48$ & $5.58$ & $13.98$ & $11.85$ &
$0.080$ & $0.074$ \\

\hline \hline

\end{tabular}
\end{table}
\endgroup

We have also performed calculations to investigate the $\beta$-In$_2$X$_2$
polytypes.  We find that these structures are dynamically stable, but the
static-lattice cohesive energy is slightly lower than the $\alpha$ structure
by 0.022 and 0.013 eV per unit cell according to the LDA and PBE functionals,
respectively.  The relative energy of the $\alpha$ and $\beta$ polytypes is
almost the same for each chalcogen X\@.  The phonon zero-point energies (ZPEs)
reported in Table \ref{table:parameters_summary} demonstrate that lattice
dynamics does not affect the relative stability of the $\alpha$ and $\beta$
polytypes.  The optimal lattice parameters of these structures are summarized
in the bottom half of Table \ref{table:parameters_summary}.

\subsection{Lattice dynamics \label{sec:phonons}}

\begin{figure*}
\begin{center}
\includegraphics[clip,scale=0.45]{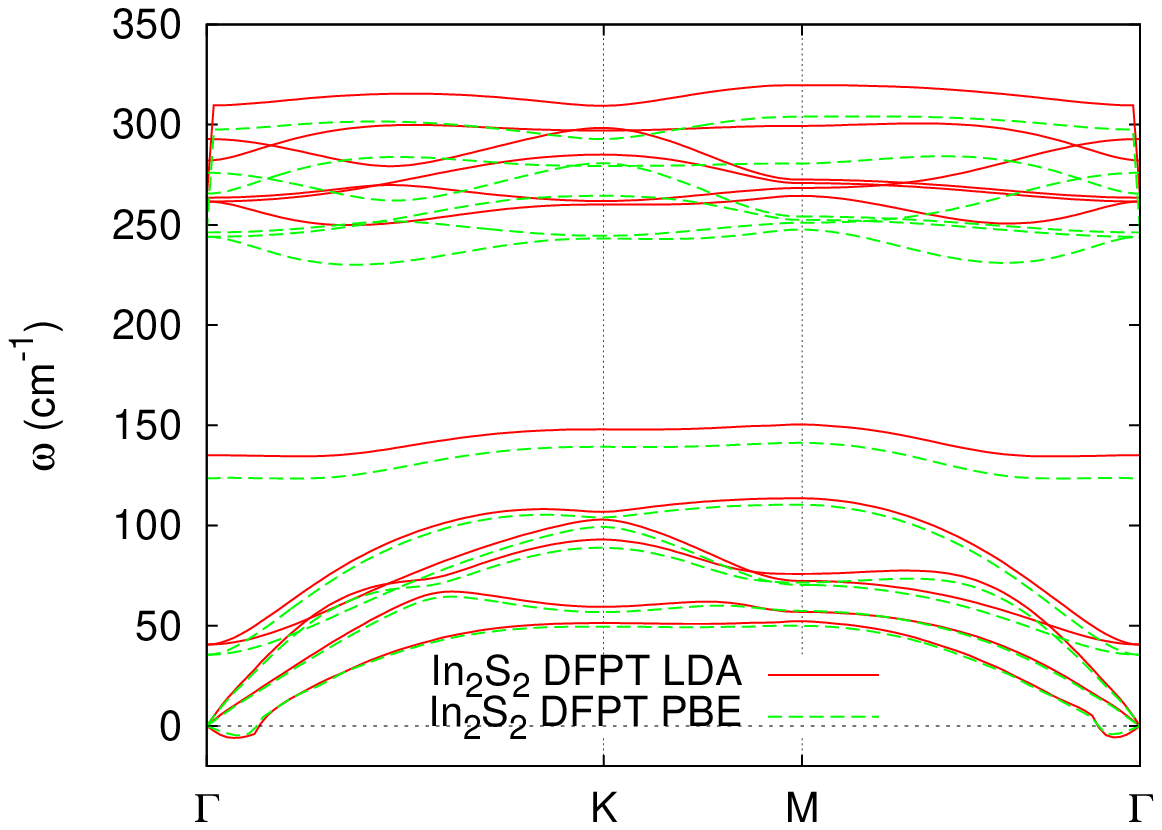}
\includegraphics[clip,scale=0.45]{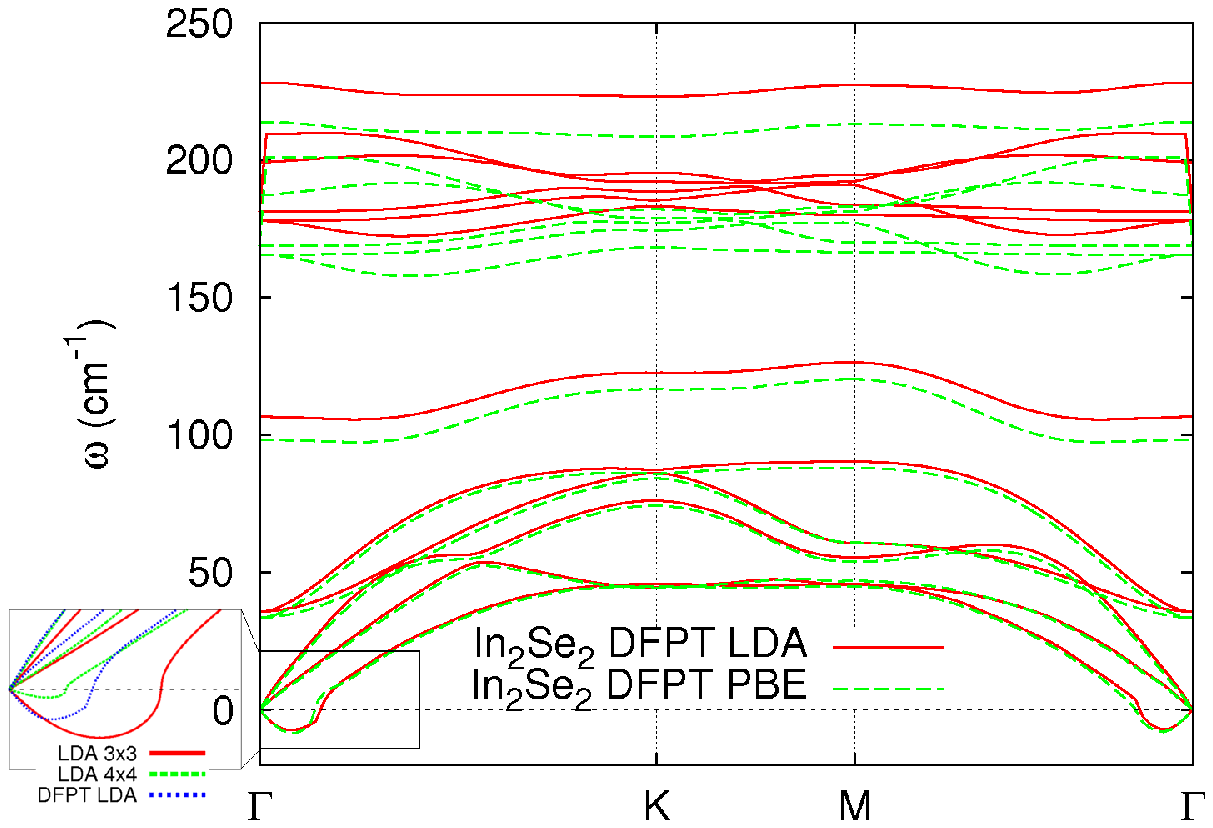}
\includegraphics[clip,scale=0.45]{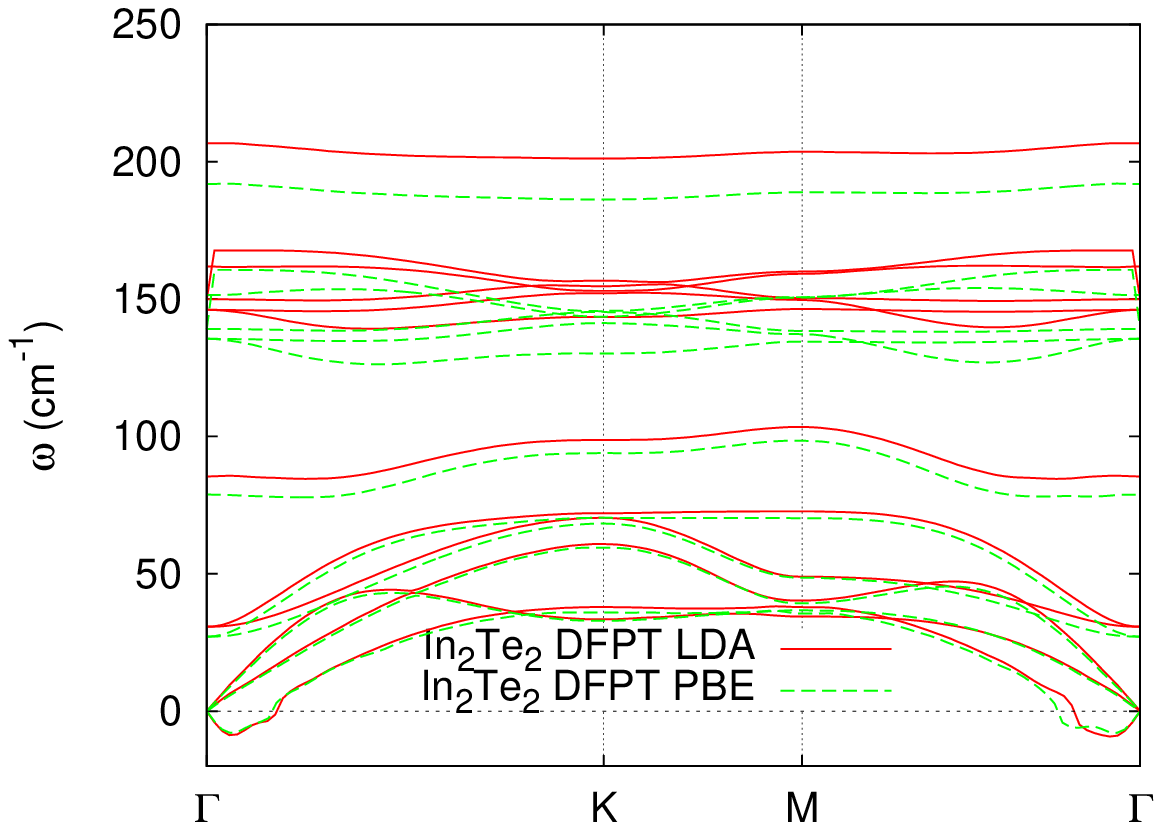}
\\\includegraphics[clip,scale=0.45]{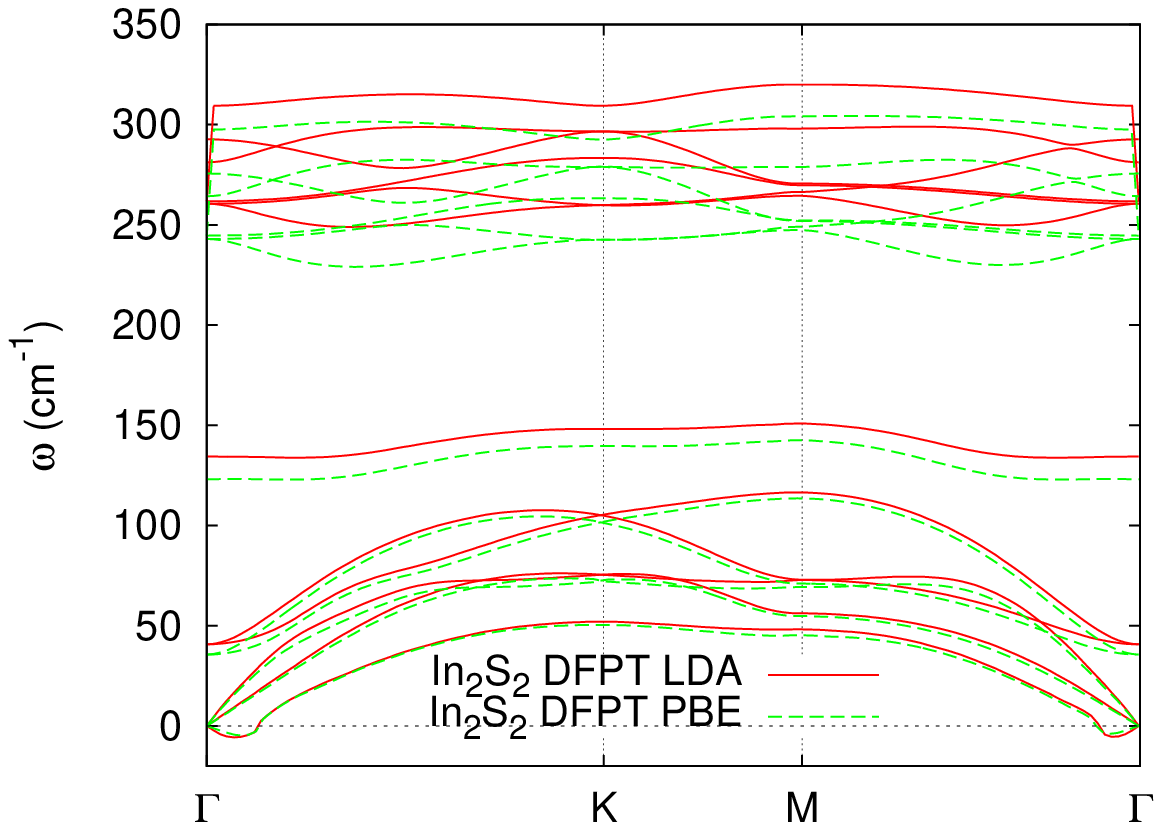}
\includegraphics[clip,scale=0.45]{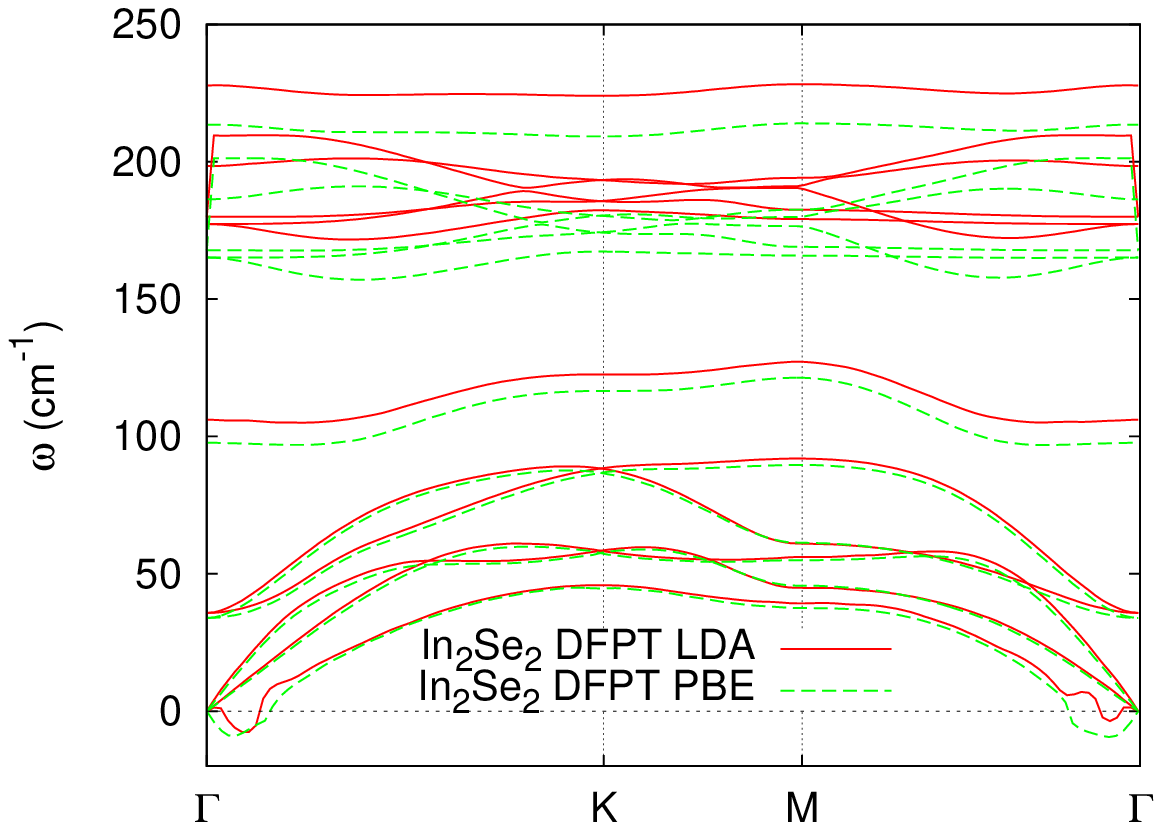}
\includegraphics[clip,scale=0.45]{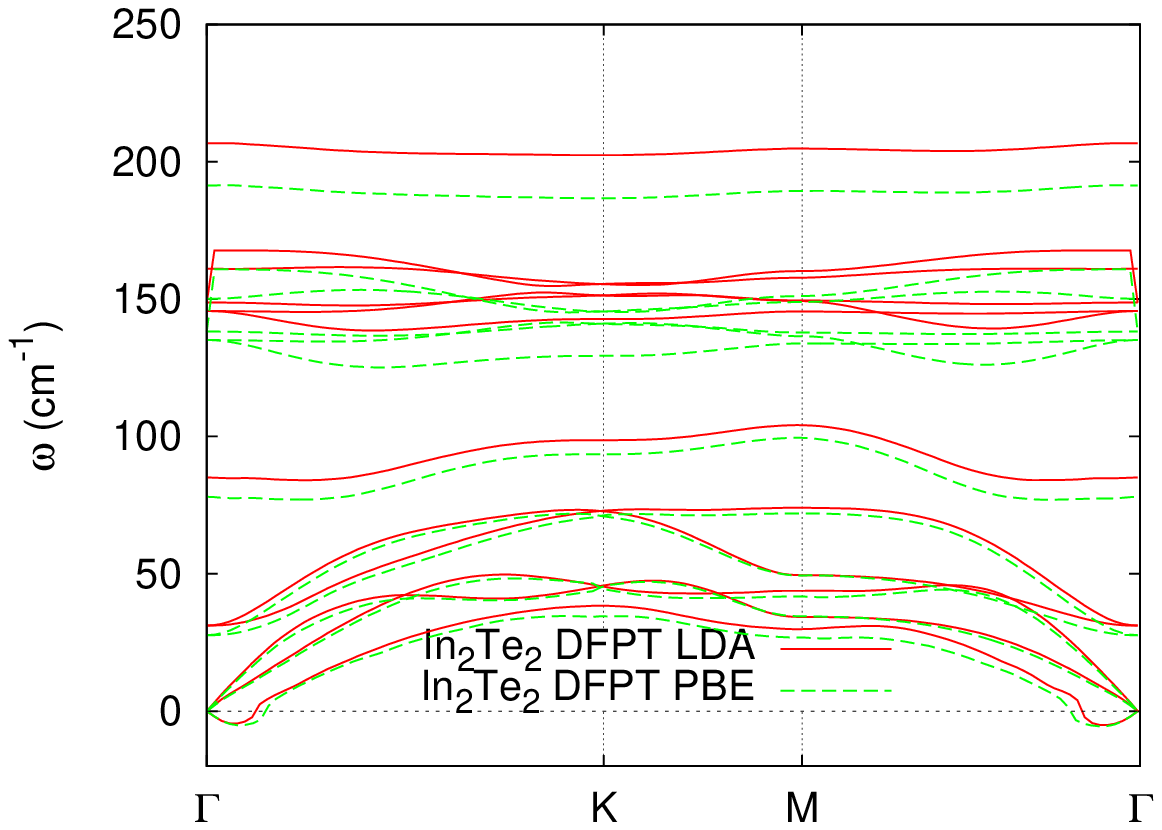}
\caption{(Color online) Phonon dispersion curves for $\alpha$ (top panel) and
  $\beta$ (bottom panel) polytypes of In$_2$S$_2$, In$_2$Se$_2$, and
  In$_2$Te$_2$. The inset shows the low-frequency spectrum of
  $\alpha$-In$_2$Se$_2$ with several methods.  Below we list the DFT-LDA
  optical-phonon frequencies at $\Gamma$, the irreducible representation
  (irrep.)\ to which the eigenvectors belong, and the IR and Raman activity.
  The modes are labeled as longitudinal optical (LO), transverse optical (TO),
  or out-of-plane optical (ZO)\@. The irreducible representation is given in
  the conventional molecular notation in which one and two primes indicate $z
  \rightarrow -z$ reflection symmetry and antisymmetry, respectively. For IR
  activity we indicate the component of electric field involved (out-of-plane,
  $E_z$, or in-plane, $E_\|$), while for Raman activity we indicate the
  components of electric field that are coupled by the Raman tensor.
\label{fig:phonons}}
\begin{tabular}{lccccccccccc}
\hline \hline

\multicolumn{9}{c}{$\alpha$-In$_2$X$_2$}\\ &
\multicolumn{3}{c}{$\omega_{\Gamma }$ (cm$^{-1  }$)} & & \multicolumn{3}{c}{IR
  intensity\ (D$^2${\AA}$^{-2}$amu$^{-1}$)} & Polarization of Raman- \\

\raisebox{1.5ex}[0pt]{Branch} & In$_2$S$_2$ & In$_2$Se$_2$ & In$_2$Te$_2$ &
\raisebox{1.5ex}[0pt]{Irrep.}  & In$_2$S$_2$ & In$_2$Se$_2$ & In$_2$Te$_2$ &
active modes \\

\hline

 4 & $40.6$ & $35.6$ & $30.7$ & $E^{\prime \prime}$ & -- & -- & -- & $E_z
 \leftrightarrow E_\|$ \\

 5 & $40.6$ & $35.6$ & $30.7$ & $E^{\prime \prime}$ & -- & -- & -- & $E_z
 \leftrightarrow E_\|$ \\

 6 & $135$ & $107$ & $85.4$ & $A^\prime_1$ & -- & -- & -- & $\left\{ {E_\|
   \leftrightarrow E_\| \atop E_z \leftrightarrow E_z} \right.$ \\

 7 & $262$ & $178$ & $146$ & $E^{\prime \prime}$ & -- & -- & -- & $E_z
 \leftrightarrow E_\|$ \\

 8 & $262$ & $178$ & $146$ & $E^{\prime \prime}$ & -- & -- & -- & $E_z
 \leftrightarrow E_\|$ \\

 9 (TO) & $264$ & $181$ & $150$ & $E^\prime$ & $10.2$ ($E_\|$) & 5.18 & 3.57 &
 $E_\| \leftrightarrow E_\|$ \\

10 (LO) & $264$ & $181$ & $150$ & $E^\prime$ & $10.2$ ($E_\|$) & 5.18 & 3.57 &
$E_\| \leftrightarrow E_\|$ \\

11 (ZO) & $282$ & $199$ & $162$ & $A_2^{\prime \prime}$ & $0.25$ ($E_z$) &
0.10 & 0.061 & -- \\

12 & $293$ & $228$ & $207$ & $A_1^\prime$ & -- & -- & -- & $\left\{ {E_\|
  \leftrightarrow E_\| \atop E_z \leftrightarrow E_z} \right.$ \\

\hline \hline \multicolumn{9}{c}{$\beta$-In$_2$X$_2$}\\ &
\multicolumn{3}{c}{$\omega_{\Gamma }$ (cm$^{-1 }$)} & & \multicolumn{3}{c}{IR
  intensity\ (D$^2${\AA}$^{-2}$amu$^{-1}$)} & Polarization of Raman- \\

\raisebox{1.5ex}[0pt]{Branch} & In$_2$S$_2$ & In$_2$Se$_2$ & In$_2$Te$_2$ &
\raisebox{1.5ex}[0pt]{Irrep.}  & In$_2$S$_2$ & In$_2$Se$_2$ & In$_2$Te$_2$ &
active modes \\

\hline

4 & 40.8 & 35.8 & 31.2 & $E_g$ & -- & -- & -- & $\left\{ {E_\| \leftrightarrow
  E_\| \atop E_\| \leftrightarrow E_z} \right.$ \\

5 & 40.8 & 35.8 & 31.2 & $E_g$ & -- & -- & -- & $\left\{ {E_\| \leftrightarrow
  E_\| \atop E_\| \leftrightarrow E_z} \right.$ \\

6 & 134 & 106 & 84.9 & $A_{1g}$ & -- & -- & -- & $\left\{ {E_\|
  \leftrightarrow E_\| \atop E_z \leftrightarrow E_z} \right.$ \\

7 & 261 & 177 & 146 & $E_g$ & -- & -- & -- & $\left\{ {E_\| \leftrightarrow
  E_\| \atop E_\| \leftrightarrow E_z} \right.$ \\

8 & 261 & 177 & 146 & $E_g$ & -- & -- & -- & $\left\{ {E_\| \leftrightarrow
  E_\| \atop E_\| \leftrightarrow E_z} \right.$ \\

9 (TO) & 262 & 180 & 149 & $E_u$ & 10.4 ($E_\|$) & 5.4 & 3.8 & -- \\

10 (LO) & 262 & 180 & 149 & $E_u$ & 10.4 ($E_\|$) & 5.4 & 3.8 & -- \\

11 (ZO) & 281 & 198 & 161 & $A_{2u}$ & 0.25 ($E_z$) & 0.10 & 0.06 & -- \\

12 & 293 & 228 & 207 & $A_{1g}$ & -- & -- & -- & $\left\{ {E_\|
  \leftrightarrow E_\| \atop E_z \leftrightarrow E_z} \right.$ \\

\hline \hline
\end{tabular}
\end{center}
\end{figure*}

We have calculated phonon dispersion curves for In$_2$X$_2$ using both the
finite-displacement approach and DFPT\@. The DFPT results are presented in
Fig.\ \ref{fig:phonons}. The finite-displacement approach agrees very well
with these dispersion curves at a supercell size of $4 \times 4$ primitive
unit cells. Other than a small pocket near $\Gamma$, we find no trace of
imaginary frequencies in the Brillouin zone. This small pocket of instability
(shown in detail in the inset beside the middle panel of
Fig.\ \ref{fig:phonons} for $\alpha$-In$_2$Se$_2$) is extremely sensitive to
the details of the calculation and in some cases disappears altogether. This
suggests that it merely indicates the difficulty of achieving numerical
convergence for the flexural phonon branch, which appears to be a common issue
in first-principles calculations for 2D materials.\cite{footnote:instability}
Therefore the phonon dispersion curves suggest that isolated atomic crystals
of hexagonal indium chalcogenides, In$_2$X$_2$, are dynamically stable.  The
spurious imaginary modes were assumed not to contribute to the ZPEs reported
in Table \ref{table:parameters_summary}.  The nonanalytic contribution to the
dynamical matrix due to long-range Coulomb interactions
(longitudinal/transverse optical mode splitting) is neglected in this work.
For a discussion of this issue in 2D materials, see App.\ A of
Ref.\ \onlinecite{sanchez-portal}.

The DFT-LDA phonon dispersions for $\alpha$- and $\beta$-In$_2$X$_2$ are shown
in Fig.\ \ref{fig:phonons}. The caption for this figure also contains a
tabulated list of all IR- and Raman-active optical phonon modes at the
$\Gamma$ point.  We have used a unit cell with lattice vectors
$(a/2,\sqrt{3}a/2)$ and $(-a/2,\sqrt{3}a/2)$, where $a$ is the lattice
parameter.  The lattice parameters and other structural parameters are given
in a separate Table \ref{table:parameters_summary}.  $\hat{\bf x}$, $\hat{\bf
  y}$, and $\hat{\bf z}$ are unit vectors in the Cartesian directions.  The
most important difference between the $\alpha$ and $\beta$ structures is the
number of Raman-active $\Gamma$-point phonons. We find that there are two
fewer Raman-active modes in $\beta$-In$_2$X$_2$, offering a way to distinguish
the polytypes.  Note that $\beta$-In$_2$X$_2$ possesses inversion symmetry,
while $\alpha$-In$_2$X$_2$ does not.  Raman and IR activity are mutually
exclusive in materials with inversion symmetry.  If none of the IR-active
modes found in In$_2$X$_2$ appears in the Raman spectrum of a sample, this
would point towards the $\beta$-In$_2$X$_2$ polytype.  We discuss the
electronic band structure of the energetically more favorable $\alpha$ phase
in Sec.\ \ref{sec:alpha}, and then discuss the $\beta$ phase in
Sec.\ \ref{sec:beta}.

\section{Electronic and optical properties of monolayers of
  $\alpha$-In$_2$X$_2$  \label{sec:alpha}}

\subsection{Band structures}

\begin{figure*}
\begin{center}
\includegraphics[clip,scale=0.45]{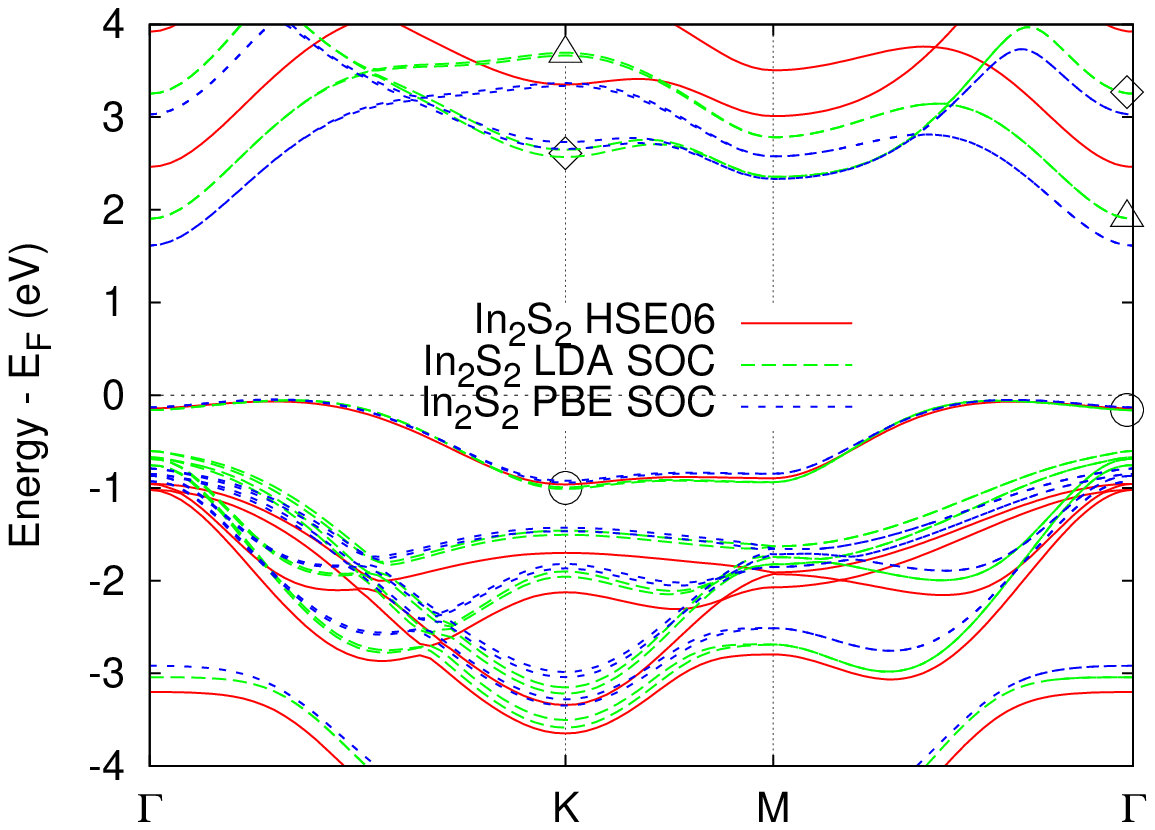}
\includegraphics[clip,scale=0.45]{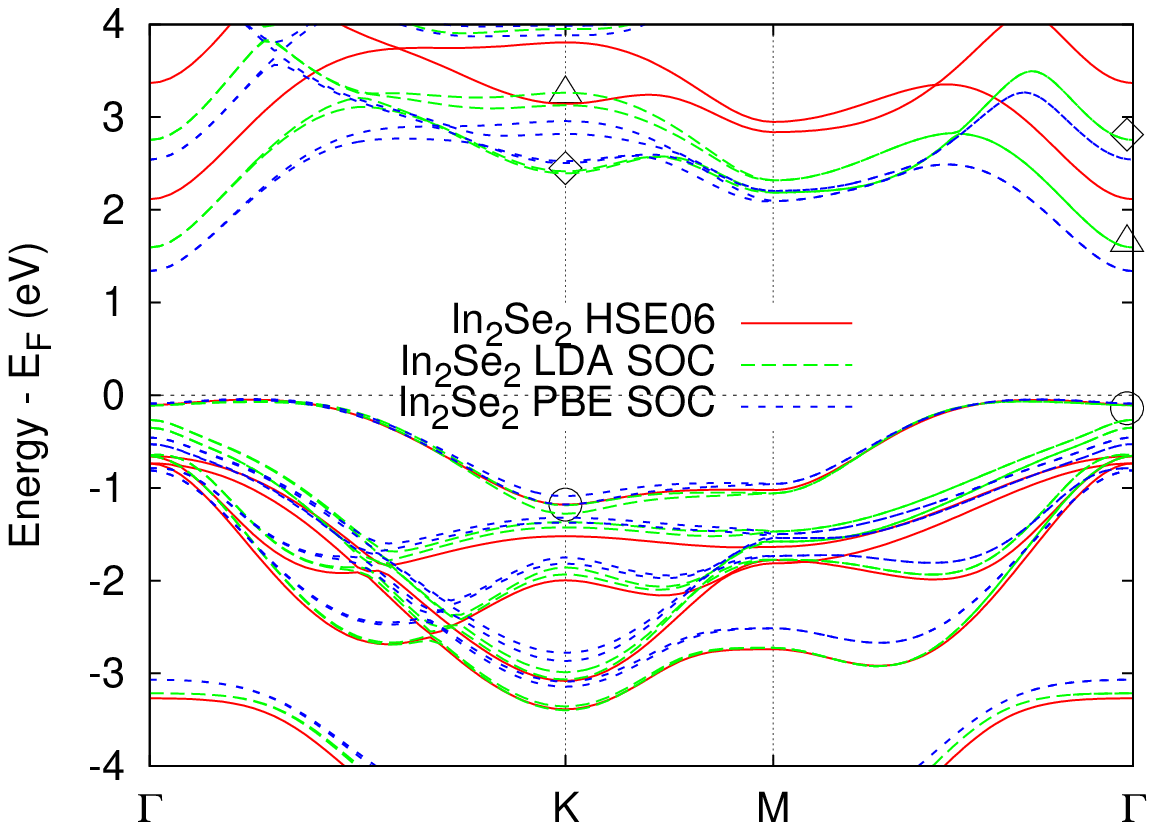}
\includegraphics[clip,scale=0.45]{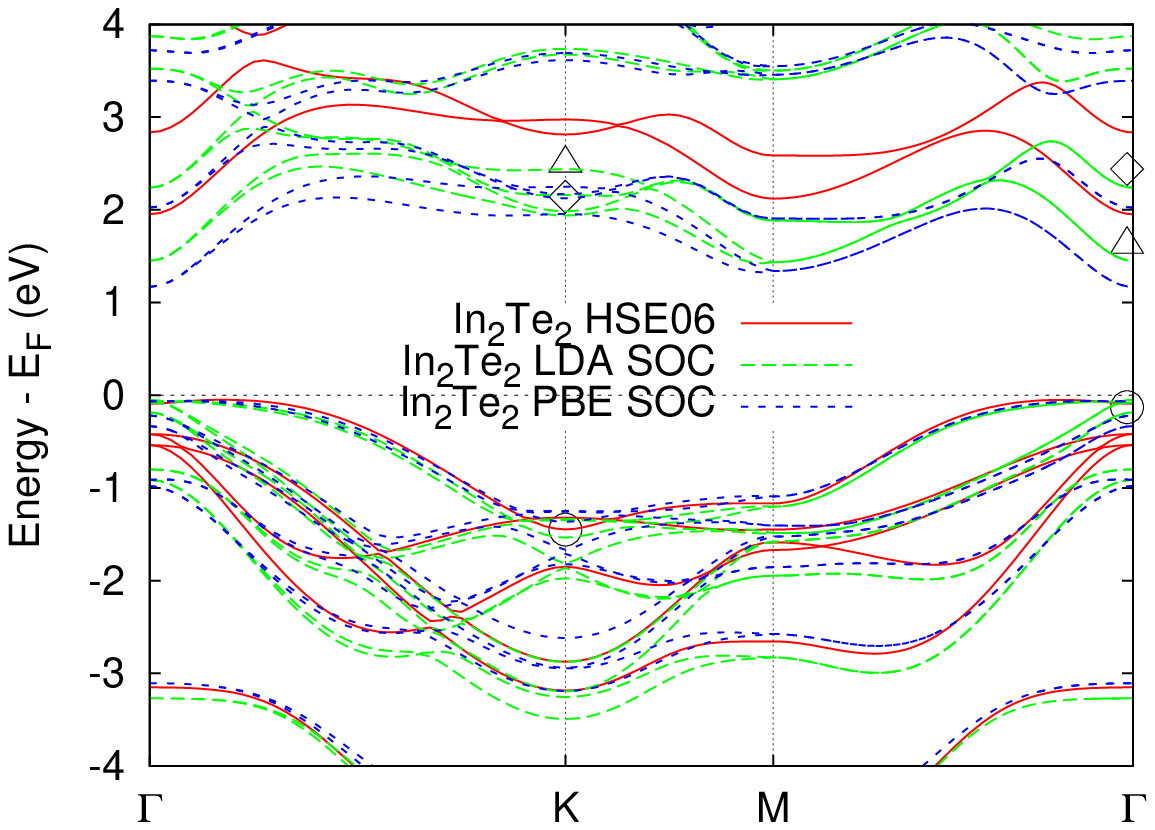}
\caption{(Color online) HSE06 band structures (solid red lines) for
  $\alpha$-In$_2$S$_2$, $\alpha$-In$_2$Se$_2$, and $\alpha$-In$_2$Te$_2$ (top
  panel). Spin--orbit coupling (SOC) is not included in these results. The
  zero of energy is taken to be the Fermi level $E_F$ and the bottom of the
  conduction band is marked with a horizontal line. For comparison, the
  semilocal band structures are also shown, including the effects of SOC\@.
  The orbital composition of the $\alpha$-In$_2$X$_2$ states highlighted by
  $\bigcirc$, $\triangle$, and $\diamondsuit$ are summarized in the table
  below.  Dominant contributions were found to originate from $s$- and
  $p$-type orbitals; the ``$+$'' and ``$-$'' subscripts refer to even ($+$)
  and odd ($-$) states with respect to $z\rightarrow -z$ reflection. The LDA
  spin-orbit splittings $|\Delta E_{\rm SO}^\textrm{K}|$ of the bands at the K
  point are also given.  The notation ``$p_x p_y$'' refers to equal $p_x$ and
  $p_y$ contributions as a consequence of symmetry.  \label{fig:bands}}
\begin{tabular}{lccccc}
\hline \hline

X & Band & $\Gamma$ && K &\quad $|\Delta E_{\rm SO}^\textrm{K}|$ (meV) \\

\hline

S & $\bigcirc_{+}$& $0.012s^{\rm In}+0.039p_z^{\rm In}+0.002s^{\rm
  S}+0.198p_z^{\rm S}$ && $0.061s^{\rm In}+0.142p_z^{\rm In}+0.045p_x^{\rm
  S}p_y^{\rm S}$ & $18$ \\

S & $\triangle_{-}$& $0.127s^{\rm In}+0.003p_z^{\rm In}+0.068s^{\rm
  S}+0.081p_z^{\rm S}$ && $0.202s^{\rm In}+0.008p_z^{\rm In}+0.057p_x^{\rm
  S}p_y^{\rm S}$ & \\

S & $\diamondsuit_{+}$& $0.059s^{\rm In}+0.112p_z^{\rm In}+0.071s^{\rm
  S}+0.001p_z^{\rm S}$ && $0.028p_x^{\rm In}p_y^{\rm In}+0.037p_x^{\rm
  S}p_y^{\rm S}$ & $79$ \\

\hline

Se & $\bigcirc_{+}$& $0.011s^{\rm In}+0.044p_z^{\rm In}+0.001s^{\rm
  Se}+0.197p_z^{\rm Se}$ && $0.052s^{\rm In}+0.138p_z^{\rm In}+0.049p_x^{\rm
  Se}p_y^{\rm Se}$ & $92$  \\

Se & $\triangle_{-}$& $0.115s^{\rm In}+0.005p_z^{\rm In}+0.060s^{\rm
  Se}+0.090p_z^{\rm Se}$ && $0.193s^{\rm In}+0.007p_z^{\rm In}+0.058p_x^{\rm
  Se}p_y^{\rm Se}$ & \\

Se & $\diamondsuit_{+}$& $0.056s^{\rm In}+0.116p_z^{\rm In}+0.065s^{\rm
  Se}+0.001p_z^{\rm Se}$ && $0.028p_x^{\rm In}p_y^{\rm In}+0.036p_x^{\rm
  Se}p_y^{\rm Se}$ & $23$ \\

\hline

Te & $\bigcirc_{+}$& $0.013s^{\rm In}+0.053p_z^{\rm In}+0.001s^{\rm
  Te}+0.168p_z^{\rm Te}$ && $0.039s^{\rm In}+0.131p_z^{\rm In}+0.047p_x^{\rm
  Te}p_y^{\rm Te}$ & $13$ \\

Te & $\triangle_{-}$& $0.119s^{\rm In}+0.007p_z^{\rm In}+0.067s^{\rm
  Te}+0.079p_z^{\rm Te}$ && $0.167s^{\rm In}+0.005p_z^{\rm In}+0.052p_x^{\rm
  Te}p_y^{\rm Te}$ & \\

Te & $\diamondsuit_{+}$& $0.064s^{\rm In}+0.103p_z^{\rm In}+0.063s^{\rm
  Te}+0.004p_z^{\rm Te}$ && $0.029p_x^{\rm In}p_y^{\rm In}+0.030p_x^{\rm
  Te}p_y^{\rm Te}$ & $47$ \\

\hline \hline
\end{tabular}
\end{center}
\end{figure*}

\begin{figure}
\begin{center}
\includegraphics[clip,scale=0.6]{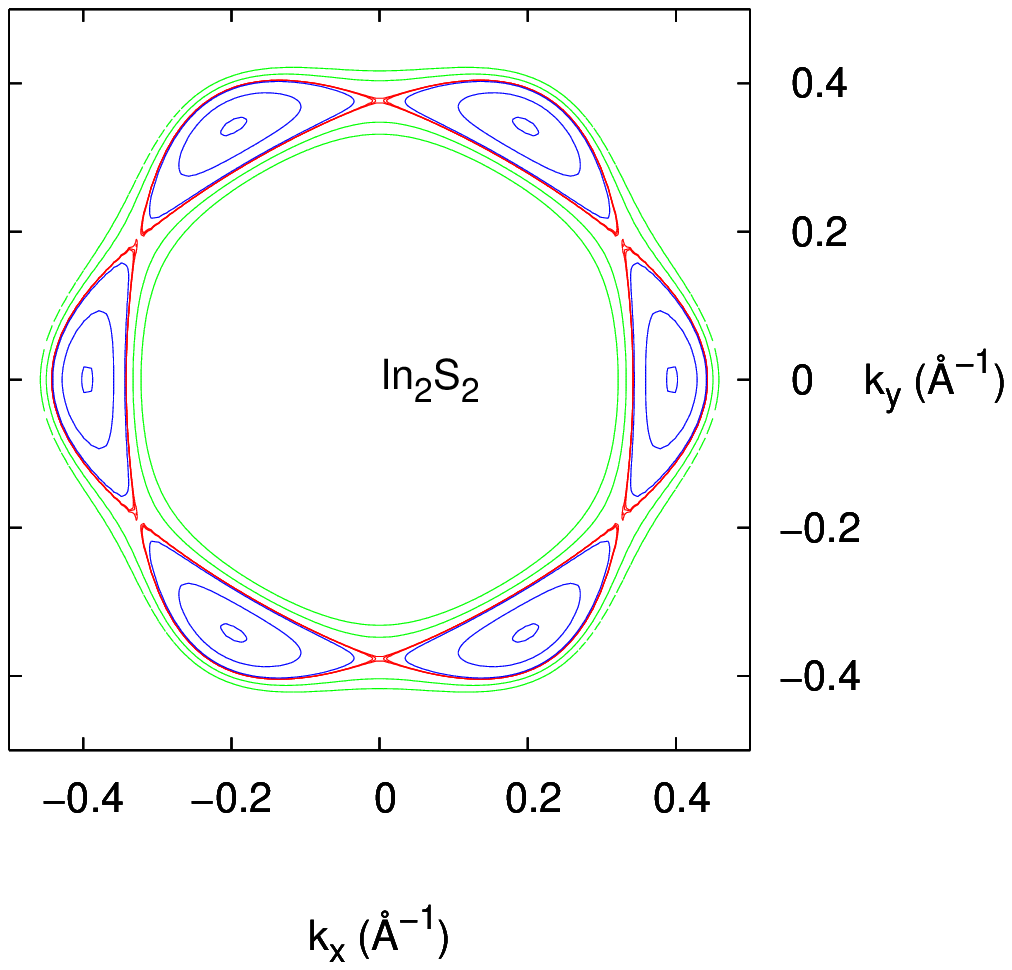}
\\ \includegraphics[clip,scale=0.6]{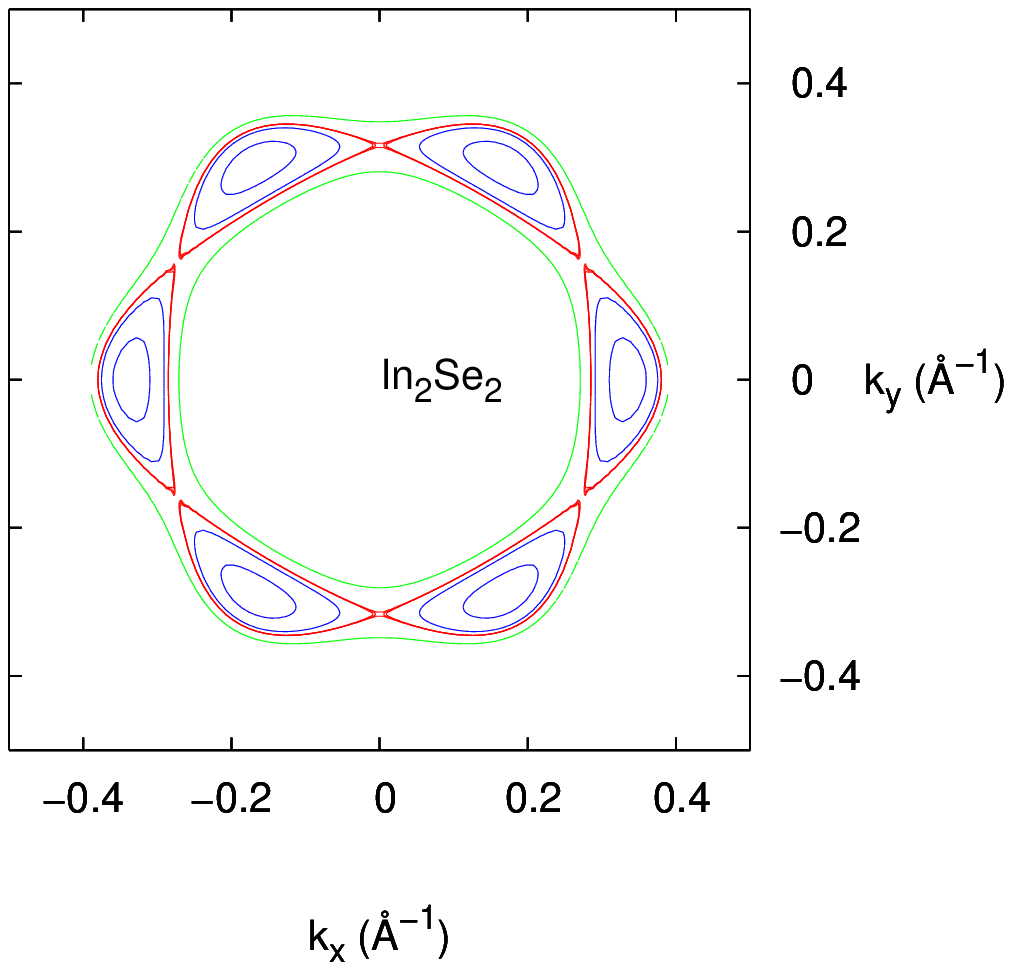}
\\ \includegraphics[clip,scale=0.6]{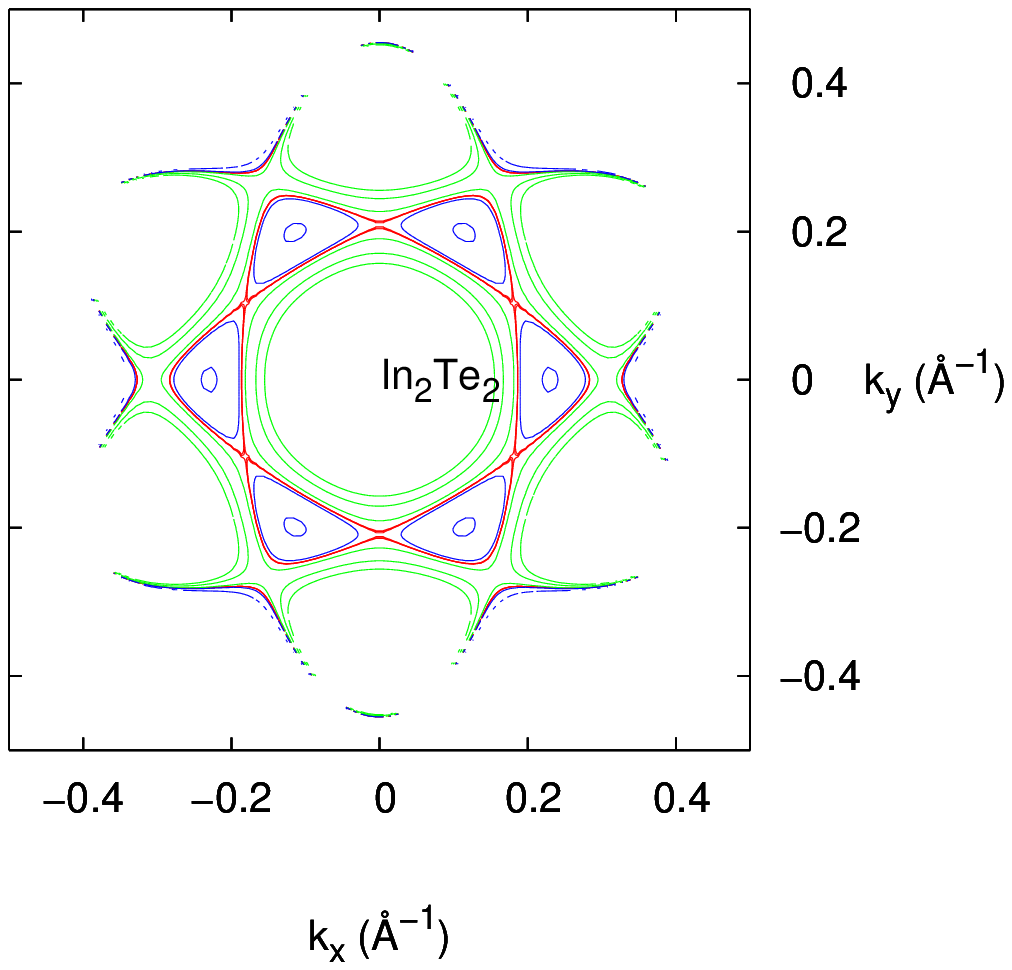}
\caption{(Color online) LDA energy contours (with a step of 2 meV) for the
  valence band of $\alpha$-In$_2$X$_2$ around the $\Gamma$-point. The contour
  corresponding to the energy of the saddle point (Lifshitz transition) is
  highlighted.  The table below shows the fitted coefficients E$_{2i}$ (in
  units of eV{\AA}$^{2i}$) for the inverted sombrero dispersion near the VBM
  of $\alpha$-In$_2$X$_2$ in Eq.\ (\ref{eq:one}).  The zero of energy is set
  to the VBM\@.  The root mean square of the residuals $\sigma$ indicates the
  amount by which the fit is in error. The last column shows the hole density
  $n_{\rm X}$ for the Lifshitz transition.
\label{fig:contours}}
\begin{tabular}{lccccccc}
\hline \hline

X & $E_0$ & $E_2$ & $E_4$ & $E_6$ & $E^\prime_6$ & $\sigma$ (meV) & $n_{\rm
  X}$ ($10^{13}$ cm$^{-2}$) \\

\hline

S  & $-0.16$ & $0.96$ & $-3.33$ & $0.42$ & $0.67$ & $0.12$ &$6.86$ \\

Se & $-0.14$ & $0.91$ & $-4.23$ & $-0.60$& $1.64$ & $0.17$ &$6.20$ \\

Te & $-0.13$ & $1.42$ & $-20.8$ & $82.3$ & $11.5$ & $0.25$ &$2.86$ \\

\hline \hline
\end{tabular}
\end{center}
\end{figure}

The calculated electronic band structures of $\alpha$-In$_2$X$_2$ are
summarized in Fig.\ \ref{fig:bands}, with the orbital compositions and
spin--orbit splittings tabulated in the figure caption.  All three materials
are indirect-gap semiconductors, primarily due to the valence-band maximum
(VBM) lying between the $\Gamma$ and K points.  Further analysis of the
valence band reveals a saddle point along the $\Gamma$--M line, illustrated in
Fig.\ \ref{fig:contours}.  This saddle point gives rise to a Van Hove
singularity in the density of states. Due to the presence of these saddle
points, hole-doping causes In$_2$X$_2$ to undergo a Lifshitz transition when
the hole concentration reaches the critical value where all states are
depleted above the energy of the saddle point, since this leads to a change in
the topology of the Fermi surface. The carrier density at which the Lifshitz
transition takes place in each material is tabulated in the caption of
Fig.\ \ref{fig:contours} and was obtained by integrating the DFT density of
states from the saddle point to the valence-band edge.

It is possible to fit an inverted sombrero polynomial to the valence-band
dispersions $E_{\rm VB}$ around the VBM:
\begin{eqnarray}
\label{eq:one}
E_{\rm VB} &=& \sum_{i=0}^{3}{E_{2i}k^{2i}} + {E^{\prime}_{6}k^{6}}\cos(6
\varphi),
\end{eqnarray}
where $k$ and $\varphi$ are the radial and polar coordinates of wave vectors
about the $\Gamma$ point.  The polar angle $\varphi$ is measured from the
$\Gamma$--K line.  The parameters $\{E_{2i}\}$ and $E^\prime_6$ were obtained
by fitting Eq.\ (\ref{eq:one}) to the DFT valence band in the ranges $0.28$
{\AA}$^{-1}<|\textbf{k}|<0.42$ {\AA}$^{-1}$, $0.22$
{\AA}$^{-1}<|\textbf{k}|<0.36$ {\AA}$^{-1}$, and $0.12$
{\AA}$^{-1}<|\textbf{k}|<0.26$ {\AA}$^{-1}$ in In$_2$S$_2$, In$_2$Se$_2$, and
In$_2$Te$_2$, respectively. The fitting ranges are centered on the position of
the VBM and their widths are chosen to ensure a quantitatively accurate fit at
both the VBM and the saddle point.  The coefficients are tabulated in the
caption of Fig.\ \ref{fig:contours}. This fit should provide a good starting
point for a simple analytical model of the valence band in these
materials. Note, however, that the fit is designed to describe the immediate
vicinity of the VBM and the saddle point, and is of limited accuracy at the
$\Gamma$ point; this is due to the fact that the quality of the fit would drop
significantly if we were to extend the fitting range as far as the $\Gamma$
point. The fitting was performed using the same procedure as that used in
Ref.\ \onlinecite{ZolyomiGaX}.

We find that the conduction-band minimum (CBM) is at the $\Gamma$ point in all
cases except the LDA band structure of $\alpha$-In$_2$Te$_2$, where it is at
the M point. The HSE06 band structure is expected to be the most reliable and
hence we predict that the CBM occurs at $\Gamma$ in all cases.  Nevertheless,
there are local minima of the conduction band at $\Gamma$, K, and M in each
case, with the exception of the PBE band structure of $\alpha$-In$_2$Te$_2$.
The HSE06 band gaps of $\alpha$-In$_2$X$_2$ are summarized in Table
\ref{table:parameters_summary2a}.  The HSE06 band gap is expected to
underestimate the quasiparticle band gap by no more than
10\%,\cite{hse10percent} and is known to be accurate in 2D
materials.\cite{hse10scuseria} The effective masses at the high-symmetry
points in the conduction band are summarized in Table
\ref{table:parameters_summary2a}. The effective mass is isotropic at the
$\Gamma$ and K points, but not at M\@. We note that the effective mass is
quite sensitive to the fitting range.  The data in Table
\ref{table:parameters_summary2a} were obtained by fitting in one dimension in
a range corresponding to $1/8$ of the K--M line in the Brillouin
zone.\cite{footnote:mass_fitting} If the fitting range is doubled, the
effective masses change by up to 10\%.

\begin{table}
\caption{HSE06 band gaps $\Delta$ and effective masses $m^\ast$ of In$_2$X$_2$
  at the high-symmetry points in the conduction band according to the HSE06
  functional (in units of electron mass $m_e$).
\label{table:parameters_summary2a}}
\begin{tabular}{lccccc}
\hline \hline

& & \multicolumn{4}{c}{$m^{\ast }/m_e$} \\

\raisebox{1.5ex}[0pt]{X} & \raisebox{1.5ex}[0pt]{$\Delta$ (eV)} & $\Gamma^c$ &
K$^c$ & M$^c_{\rightarrow \Gamma^c}$ & M$^c_{\rightarrow \textrm{K}^c}$ \\

\hline

\multicolumn{6}{c}{$\alpha$-In$_2$X$_2$} \\

\hline

S  & 2.53 & $0.26$ & $0.86$ & $1.24$ & $0.42$ \\

Se & 2.16 & $0.20$ & $0.71$ & $2.30$ & $0.33$ \\

Te & 2.00 & $0.17$ & $0.53$ & $0.64$ & $0.23$ \\

\hline

\multicolumn{6}{c}{$\beta$-In$_2$X$_2$} \\

\hline

S  & 2.45 & $0.25$ & -- & $1.59$ & $0.39$ \\

Se & 2.07 & $0.20$ & -- & $2.39$ & $0.24$ \\

Te & 1.88 & $0.16$ & -- & $0.67$ & $0.23$ \\

\hline \hline
\end{tabular}
\end{table}

The band structures computed using semilocal density functionals are also
plotted in Fig.\ \ref{fig:bands} for comparison. The LDA and PBE functionals
give very similar results to the HSE06 functional up to the Fermi level, but
above that significant discrepancies arise. This is most notable in the case
of $\alpha$-In$_2$Te$_2$, where the position of the CBM is ambiguous: the LDA
predicts that the CBM is at the M point, while the PBE functional puts it at
the $\Gamma$ point, in agreement with HSE06.  A similar behavior was found in
2D hexagonal gallium chalcogenides.\cite{ZolyomiGaX}

In the semilocal DFT calculations we took spin-orbit (SO) coupling into
account using a relativistic DFT approach.\cite{vasp} As can be seen in
Fig.\ \ref{fig:bands} (also listed in its caption), some of the bands exhibit
spin splitting, including the highest valence ($\Delta E_{\rm
  SO}^{v,\textrm{K}}$) and lowest conduction ($\Delta E_{\rm
  SO}^{c,\textrm{K}}$) bands near the K point (see the table in
Fig.\ \ref{fig:bands}). While we were unable to calculate the SO splittings in
HSE06 due to limited computational resources, we expect that they will exhibit
a similar magnitude to that found in the semilocal band structures.  The
caption of Fig.\ \ref{fig:bands} also contains lists describing the orbital
decomposition of the valence and conduction band states at the $\Gamma$ and K
points into the most relevant atomic orbitals of In and the chalcogens.

\subsection{Optical absorption spectra}

The orbital composition of the bands was obtained by projecting the orbitals
in the plane-wave basis set of \textsc{vasp} onto spherical harmonics, and the
results are reported in the caption of Fig.\ \ref{fig:bands}. We have found
that these bands around the Fermi level are dominated by $s$- and $p$-type
orbitals. Although one expects the $d$ orbitals to substantially influence the
electronic structure in In-based compounds, the valence and conduction bands
of In$_2$X$_2$ monolayers do not appear to contain any significant
contributions from $d$ states, despite the explicit inclusion of all the $d$
electrons in our calculations.  States in each band are either odd or even
with respect to $z\rightarrow -z$ symmetry (this information is obtained from
the complex phases of the orbital decomposition in \textsc{vasp}). Therefore,
the interband absorption selection rules require that photons polarized in the
plane of the 2D crystal are absorbed by transitions between bands whose wave
functions have the same $z\rightarrow -z$ symmetry (even$\rightarrow$even and
odd$\rightarrow$odd), and photons polarized along the $z$ axis cause
transitions between bands with opposite symmetry (even$\rightarrow$odd and
odd$\rightarrow$even).

\begin{figure*}
\begin{center}
\includegraphics[clip,scale=0.45]{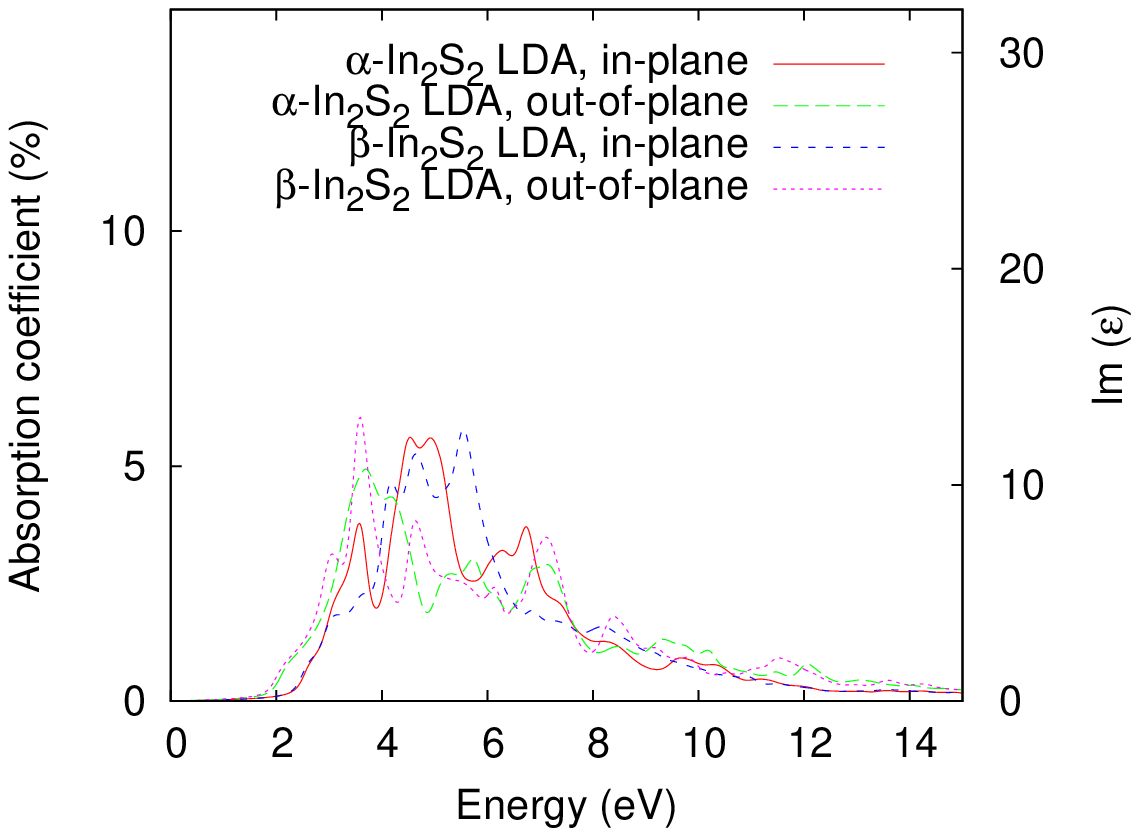}
\includegraphics[clip,scale=0.45]{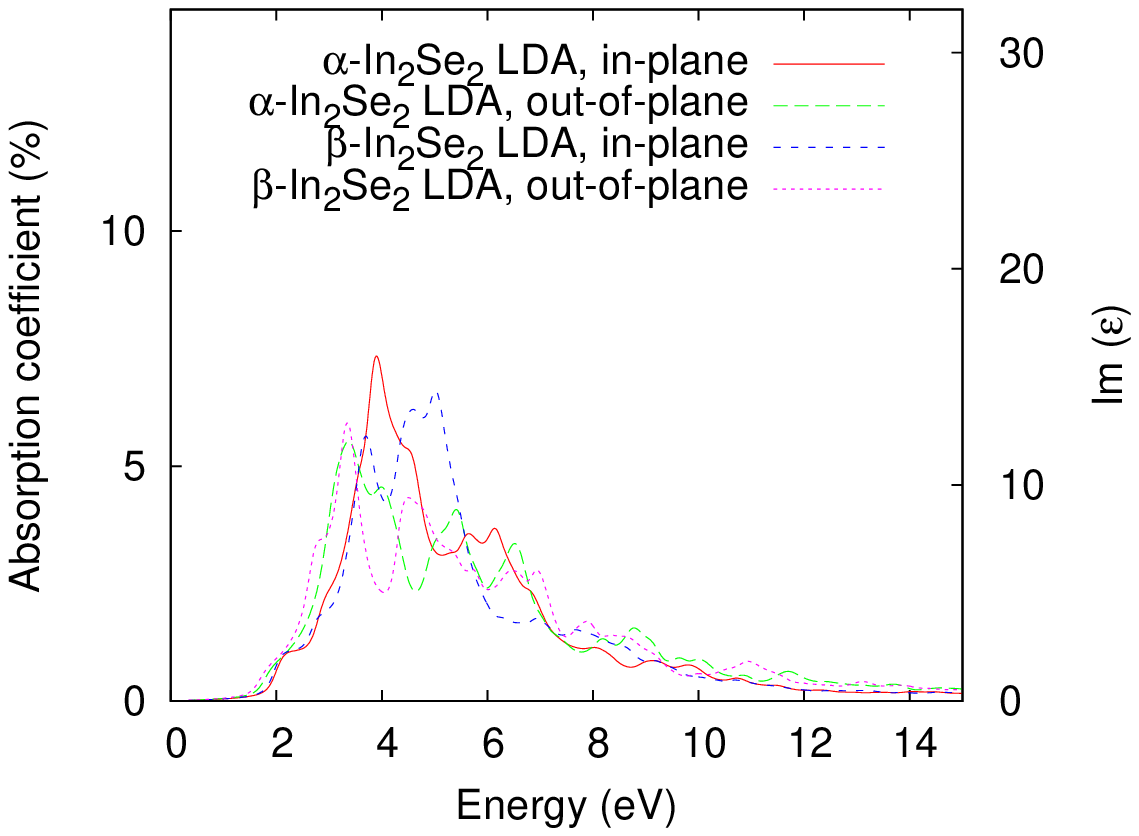}
\includegraphics[clip,scale=0.45]{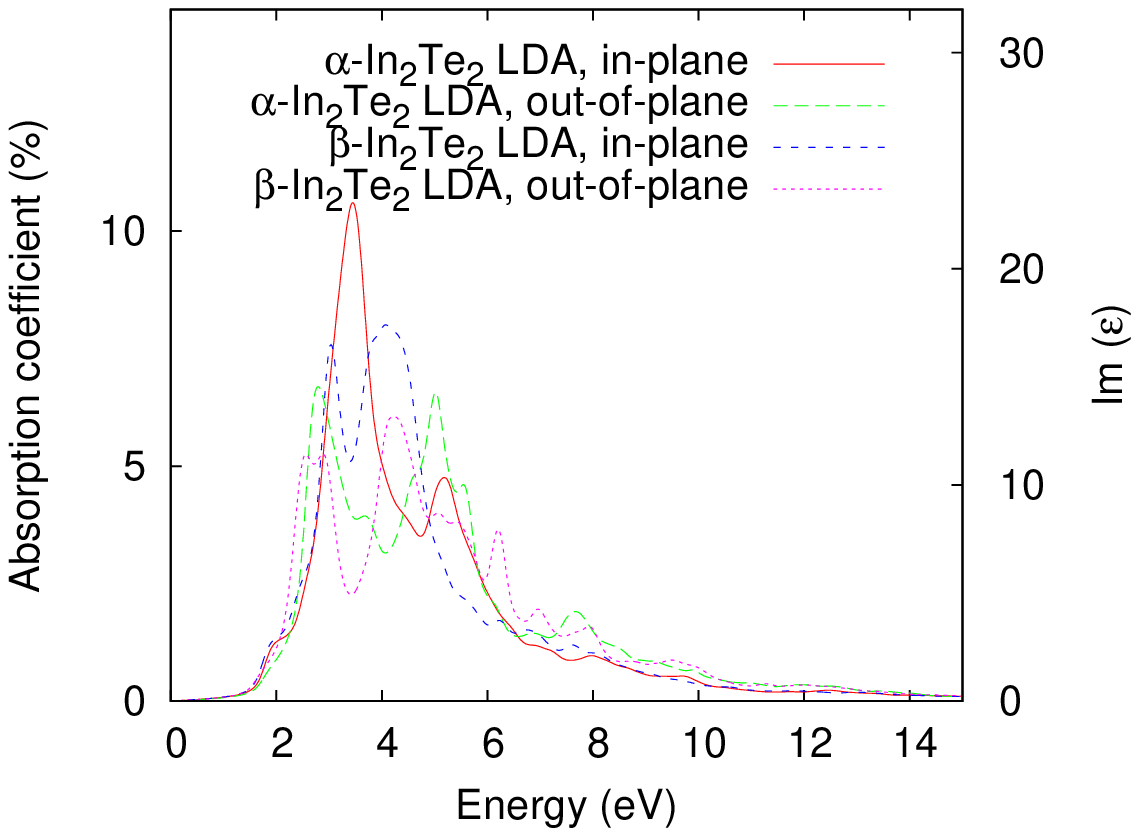}
\caption{(Color online) Absorption coefficient of $\alpha$- and
  $\beta$-In$_2$X$_2$ 2D crystals as obtained from the imaginary part of the
  dielectric function $\varepsilon$ by normalizing it to absolute units after
  it was compared to Im$(\varepsilon)$ evaluated for graphene in the range
  0.8--1.5 eV, where monolayer graphene absorbs 2.3\% of light. The raw
  results for Im$(\varepsilon)$ are indicated on the right-hand axis.
\label{fig:absorptions}
}
\end{center}
\end{figure*}

The calculated LDA optical absorption spectra are shown in
Fig.\ \ref{fig:absorptions}.  The intensities are obtained from the imaginary
part of the dielectric function and normalized to absolute units by using
graphene as a benchmark,\cite{ZolyomiGaX} since we know that graphene absorbs
2.3\% of light intensity over a broad spectral range. We calculated the LDA
dielectric function of graphene at low energies and rescaled the absorption
coefficients to reproduce the 2.3\% absorption, then applied the same scaling
to the In$_2$X$_2$ spectra.  Note that LDA results are only qualitatively
accurate and should only be used for a comparative study of the different
In$_2$X$_2$ monolayers and for an order-of-magnitude estimate of the expected
peak positions. Furthermore, local-field effects, which are expected to
influence out-of-plane absorption, are not included.  A better description
would require a computationally much more expensive calculation using the $GW$
approximation and the Bethe--Salpeter equation for excitonic
corrections.\cite{bse_example} Much like Ga$_2$X$_2$ monolayers, In$_2$X$_2$
sheets exhibit a prominent absorption peak (originating from the vicinity of
the K point) near 3--5 eV, where the absorption coefficients of In$_2$X$_2$
are comparable to and even exceed that of monolayer and bilayer graphene. As
such, we suggest that ultrathin films of InX biased in vertical tunneling
transistors with graphene electrodes could be used as an active element for
the detection of ultraviolet photons. It is not surprising to find absorptions
of a similar order of magnitude in In$_2$X$_2$ and graphene, since both are
atomically thin materials.

\section{Electronic and optical properties of monolayers of
  $\beta$-In$_2$X$_2$  \label{sec:beta}}

\subsection{Band structures}

\begin{figure*}
\begin{center}
\includegraphics[clip,width=0.31\textwidth]{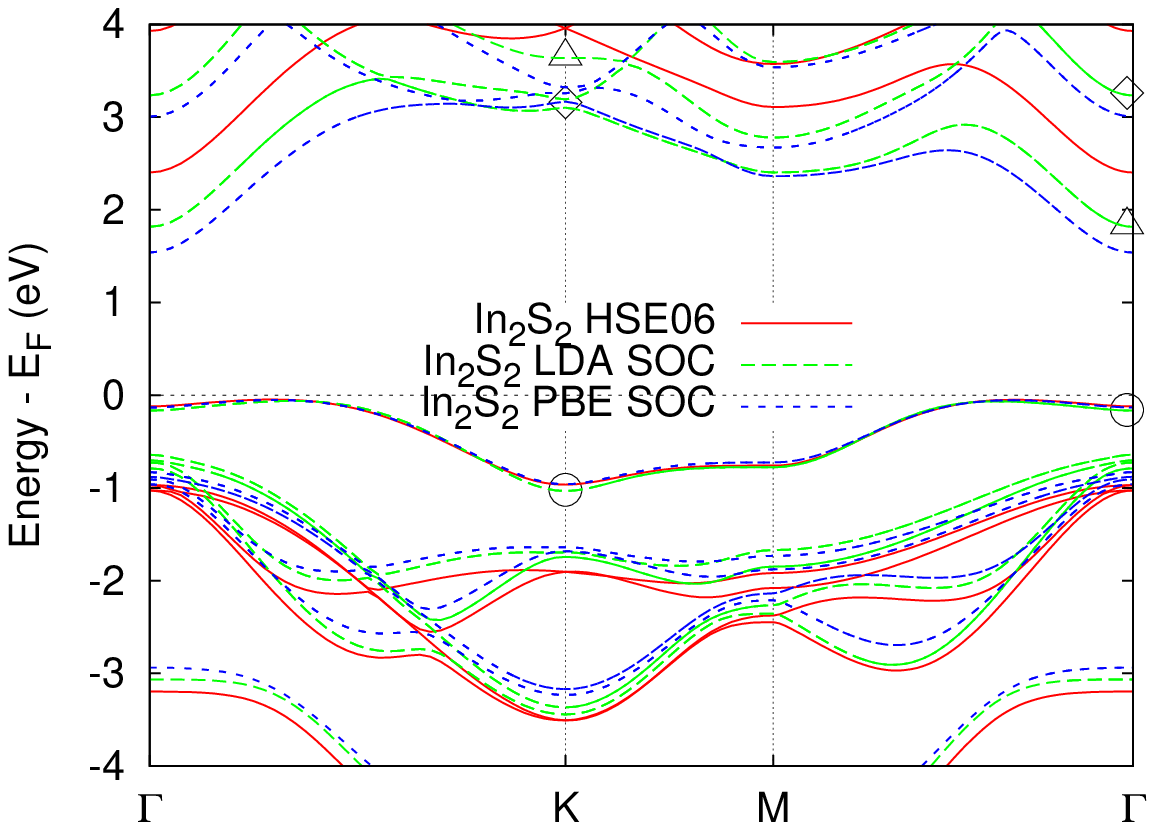}
\includegraphics[clip,width=0.31\textwidth]{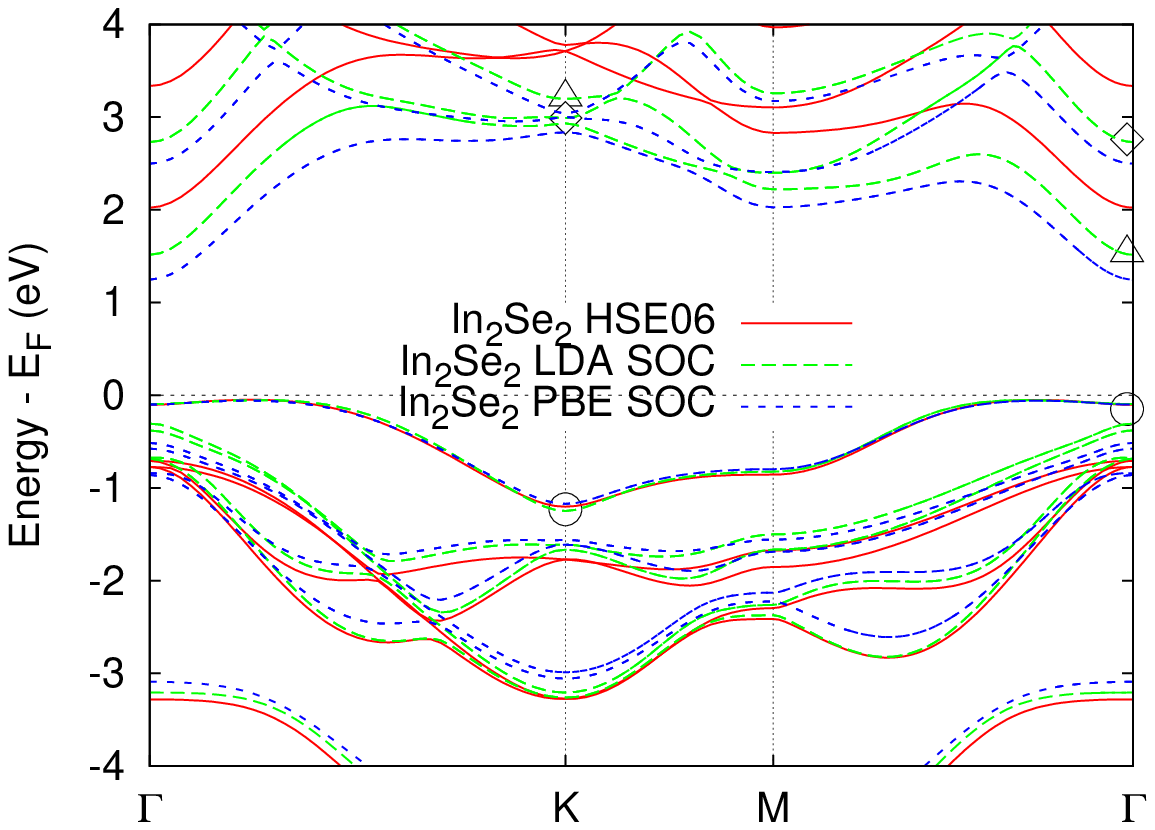}
\includegraphics[clip,width=0.31\textwidth]{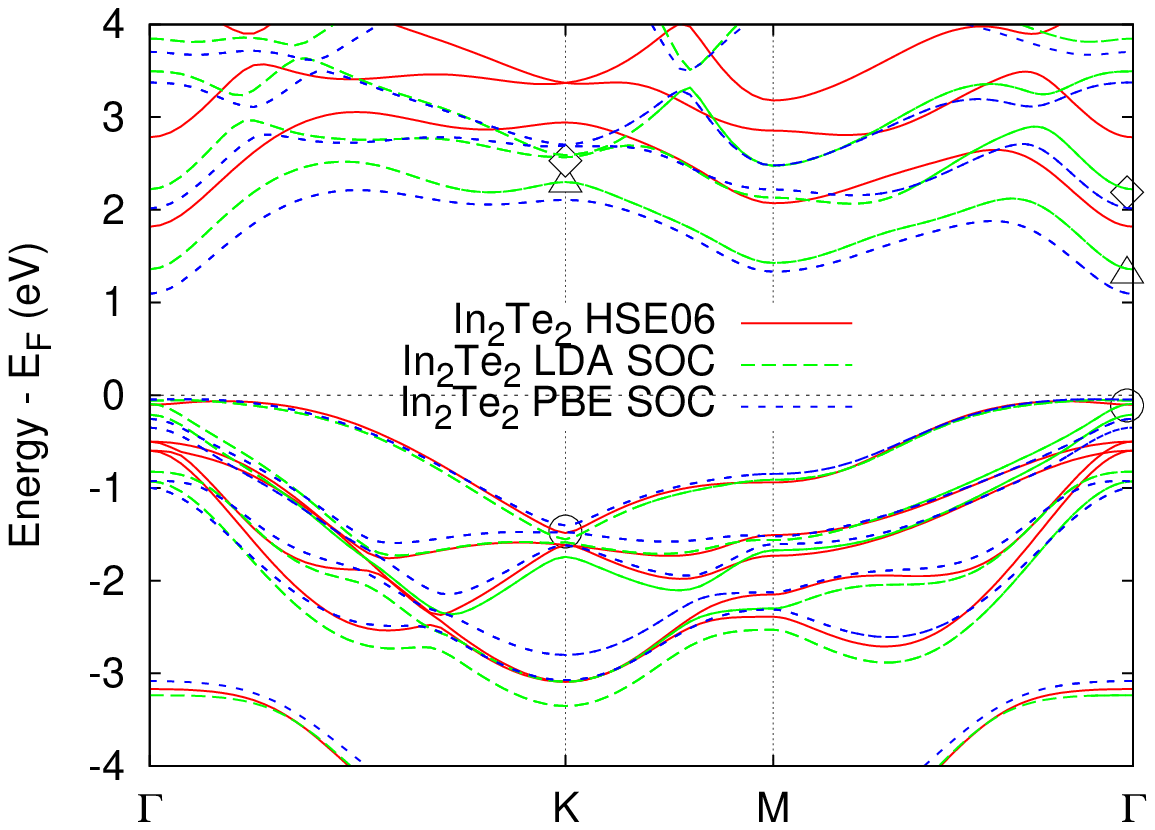}
\caption{(Color online) HSE06, LDA and PBE DFT band structures for
  $\beta$-In$_2$S$_2$, $\beta$-In$_2$Se$_2$, and
  $\beta$-In$_2$Te$_2$. Spin-orbit coupling is taken into account in the case
  of LDA and PBE\@. The zero of energy is taken to be the Fermi level $E_F$
  and the bottom of the conduction band is marked with a horizontal line.  The
  orbital composition of the $\beta$-In$_2$X$_2$ states highlighted by
  $\bigcirc$, $\triangle$, and $\diamondsuit$ are summarized in the table
  below.  Dominant contributions were found to originate from $s$- and
  $p$-type orbitals; the ``$+$'' and ``$-$'' subscripts refer to even ($+$)
  and odd ($-$) states with respect to three-dimensional inversion. The
  notation ``$p_x p_y$'' refers to equal $p_x$ and $p_y$ contributions as a
  consequence of symmetry.
\label{fig:bands_alt}
}
\begin{tabular}{lcccc}
\hline \hline

X & Band&$\Gamma$&$$&K\\

\hline

S & $\bigcirc_{+}$& $0.012s^{\rm In}+0.039p_z^{\rm In}+0.002s^{\rm
  S}+0.199p_z^{\rm S}$ && $0.060s^{\rm In}+0.142p_z^{\rm In}+0.045p_x^{\rm
  S}p_y^{\rm S}$ \\

S & $\triangle_{-}$& $0.126s^{\rm In}+0.004p_z^{\rm In}+0.067s^{\rm
  S}+0.080p_z^{\rm S}$ && $0.202s^{\rm In}+0.008p_z^{\rm In}+0.058p_x^{\rm
  S}p_y^{\rm S}$ \\

S & $\diamondsuit_{+}$& $0.060s^{\rm In}+0.112p_z^{\rm In}+0.072s^{\rm
  S}+0.001p_z^{\rm S}$ && $0.059p_x^{\rm In}p_y^{\rm In}+0.052p_x^{\rm
  S}p_y^{\rm S}+0.054p_z^{\rm S}$ \\

\hline

Se & $\bigcirc_{+}$& $0.012s^{\rm In}+0.043p_z^{\rm In}+0.001s^{\rm
  Se}+0.198p_z^{\rm Se}$ && $0.051s^{\rm In}+0.138p_z^{\rm In}+0.049p_x^{\rm
  Se}p_y^{\rm Se}$ \\

Se & $\triangle_{-}$& $0.115s^{\rm In}+0.005p_z^{\rm In}+0.059s^{\rm
  Se}+0.088p_z^{\rm Se}$ && $0.192s^{\rm In}+0.007p_z^{\rm In}+0.058p_x^{\rm
  Se}p_y^{\rm Se}$ \\

Se & $\diamondsuit_{+}$& $0.057s^{\rm In}+0.117p_z^{\rm In}+0.065s^{\rm
  Se}+0.001p_z^{\rm Se}$ && $0.060p_x^{\rm In}p_y^{\rm In}+0.049p_x^{\rm
  Se}p_y^{\rm Se}+0.061p_z^{\rm Se}$ \\

\hline

Te & $\bigcirc_{+}$& $0.014s^{\rm In}+0.053p_z^{\rm In}+0.002s^{\rm
  Te}+0.169p_z^{\rm Te}$ && $0.038s^{\rm In}+0.131p_z^{\rm In}+0.047p_x^{\rm
  Te}p_y^{\rm Te}$ \\

Te & $\triangle_{-}$& $0.117s^{\rm In}+0.008p_z^{\rm In}+0.065s^{\rm
  Te}+0.078p_z^{\rm Te}$ && $0.166s^{\rm In}+0.004p_z^{\rm In}+0.053p_x^{\rm
  Te}p_y^{\rm Te}$ \\

Te & $\diamondsuit_{+}$& $0.065s^{\rm In}+0.105p_z^{\rm In}+0.064s^{\rm
  Te}+0.004p_z^{\rm Te}$ && $0.060p_x^{\rm In}p_y^{\rm In}+0.040p_x^{\rm
  Te}p_y^{\rm Te}+0.054p_z^{\rm Te}$ \\

\hline \hline
\end{tabular}
\end{center}
\end{figure*}

Figure \ref{fig:bands_alt} depicts the electronic band structures of
$\beta$-In$_2$X$_2$, which shows that the valence band is strikingly similar
to that of the $\alpha$ structure in Fig.\ \ref{fig:bands}, with the VBM once
again between the $\Gamma$ and K points. This is due to the valence band being
dominated by the Ga orbitals, which are in the same configuration in the two
polytypes. Unsurprisingly, $\beta$-In$_2$X$_2$ possesses the same anisotropic
sombrero-shaped dispersion as $\alpha$-In$_2$X$_2$ and therefore a Lifshitz
transition can be achieved in this case as well. However, the coefficients of
the polynomial fit and the critical carrier concentration are quite different,
as shown in Table\ \ref{table:orbital_decomposition2_alt}.  The band
structures with SO coupling taken into account are also shown in
Fig.\ \ref{fig:bands_alt}, with the band wave functions decomposed into the
most relevant atomic orbitals of In and the chalcogens listed in the caption.

\begin{table}
\caption{Coefficients E$_{2i }$ (in units of eV{\AA}$^{2i}$) for the inverted
  sombrero dispersion near the VBM of $\beta$-In$_2$X$_2$ in
  Eq.\ (\ref{eq:one}) using the LDA functional.  The zero of energy is set to
  the VBM\@.  The root mean square of the residuals $\sigma$ indicates the
  amount by which the fit is in error. The last column shows the critical hole
  concentration $n_{\rm X}$ at which the Lifshitz transition takes place (see
  text).
\label{table:orbital_decomposition2_alt}}
\begin{tabular}{lccccccc}
\hline \hline

X & $E_0$ & $E_2$ & $E_4$ & $E_6$ & $E^\prime_6$ & $\sigma$ (meV) & $n_{\rm
  X}$ ($10^{13}$ cm$^{-2}$) \\

\hline

S  & $-2.26$ & $1.21$ & $-7.52$ & $10.7$ & $1.99$ & $0.17$ &$8.32$ \\

Se & $-2.32$ & $1.14$ & $-4.66$ & $3.91$& $0.76$ & $0.13$ &$6.00$ \\

Te & $-1.35$ & $1.53$ & $-23.1$ & $90.9$ & $11.1$ & $0.30$ &$8.14$ \\

\hline \hline
\end{tabular}

\end{table}

The conduction band of the $\beta$ polytype is similar to that of the $\alpha$
polytype near the $\Gamma$ point; however, some significant differences arise
at the K point, where a doubly degenerate band appears at the bottom of the
conduction band with a completely different orbital composition from the
lowest conduction band of the $\alpha$ structure. The orbital composition (see
the caption of Fig.\ \ref{fig:bands_alt}) of the valence band on the other
hand is almost identical to that found in $\alpha$-In$_2$X$_2$.

\subsection{Optical absorption spectra}

The optical absorption spectra of $\beta$-In$_2$X$_2$ are shown in
Fig.\ \ref{fig:absorptions}.  These show a good deal of similarity to those of
$\alpha$-In$_2$X$_2$.  The absorption is dominated by a large peak in the
ultraviolet range in all cases and the peak absorption exceeds that of
graphene.

\section{Conclusions \label{sec:conclusions}}

We have used DFT to show that 2D hexagonal indium chalcogenides (In$_2$X$_2$
where X is S, Se, or Te) are dynamically stable. We have identified two
polytypes of In$_2$X$_2$, and we have shown how these can be distinguished by
IR and Raman spectroscopy.  We find that all of these materials are
indirect-band-gap semiconductors with an unusual inverted-sombrero-shaped
valence band. The presence of saddle points in the valence band along the
$\Gamma$--M line leads to a Lifshitz transition in the event of hole doping,
for which we have calculated the critical carrier density. We have provided an
analytical fit of the valence-band edge and have given a qualitative
description of the optical absorption spectra, which suggest that atomically
thin films of InX could find application in ultraviolet photon detectors.

\begin{acknowledgments}
We acknowledge financial support from EC-FET European Graphene Flagship
Project, EPSRC Science and Innovation Award, ERC Synergy Grant ``Hetero2D,''
the Royal Society Wolfson Merit Award, and the Marie Curie project
CARBOTRON\@.
\end{acknowledgments}


\begin{thebibliography}{99}
\bibitem{footnote:mass_fitting} To determine the effective mass at the K
  point, we fitted ${\cal E}(k) = a_0 + a_2k^2 + a_3k^3\cos(3\phi) + a_4k^4$,
  where $k$ is the distance from the K point, $\phi$ is the polar angle and
  the $\{a_i\}$ are fitting parameters, to our energy bands along the
  K--$\Gamma$ and the K--M lines.  At the $\Gamma$ point we used a similar
  procedure, but with a fitting function ${\cal E}=a_0 + a_2k^2$.  At the M
  point, where the effective mass is anisotropic, we fitted ${\cal E}=a_0 +
  a_2k^2$ separately along the M--$\Gamma$ and M--K directions to obtain the
  effective masses along the two principal axes of the effective mass tensor.
\end{thebibliography}
\end{document}